\input amstex.tex
\documentstyle{amsppt}

\NoBlackBoxes
\topmatter
\title Gerbes, (twisted) K-theory, and the supersymmetric WZW model
\endtitle
\author Jouko Mickelsson \endauthor
\affil Mathematical Physics, Royal Institute of Technology, KTH, 
SE-106 91, Stockholm, Sweden
\endaffil
\date June 17, 2002; revised October 14, 2002 \enddate

\endtopmatter

\advance\vsize -2cm
\magnification=1200
\hfuzz=30pt

\define\tr{\text{tr}}

\define\<#1,#2>{\langle #1,#2\rangle}

\define\dep(#1,#2){\text{det}_{#1}#2}
\define\norm(#1,#2){\parallel #1\parallel_{#2}}

\document
ABSTRACT The aim of this talk is to explain how symmetry breaking 
in a quantum field theory problem leads to a study of projective
bundles, Dixmier-Douady classes, and associated gerbes. A gerbe
manifests itself in different equivalent ways. Besides the
cohomological description as a DD class, it can be defined in terms 
of a family of local line bundles or as a prolongation problem 
for an (infinite-dimensional)  principal bundle, with the fiber 
consisting of (a subgroup of) projective unitaries in a Hilbert 
space. The prolongation aspect is directly related to the appearance 
of central extensions of (broken)  symmetry groups.  We also discuss
the construction of twisted K-theory classes by families of
supercharges for the supersymmetric Wess-Zumino-Witten model.

This paper is based on a lecture at the meeting ``La 70eme Rencontre
entre Physiciens Theoriciens et Mathematiciens'' in Strasbourg, May
23-25, 2002. It is an extended version (by section 4) of a talk given 
in Fukuoka, October 3, 2001, at the meeting of the Japanese Mathematical 
Society. 



\vskip 0.5in

0. INTRODUCTION

\vskip 0.3in

In quantum field theory gerbes arise when one asks the question whether 
a given bundle of quantum mechanical projective  spaces is a projectivization
of a Hilbert space bundle. Nontrivial obstructions to the existence of
the Hilbert bundle are generated by QFT anomalies.  

An anomaly in field theory is a breakdown of a group
of (gauge) symmetries in the quantization of a classical field theory
model. A classical symmetry can be broken in the quantization of
massless fermions in a classical background field. The background 
consists typically of a curved space-time metric or a Yang-Mills 
field. Because of the breakdown of the symmetry, in the quantized 
theory one cannot identify gauge or diffeomorphism equivalent Hilbert 
spaces. The quantized 
symmetry group acts only projectivly, through 
a central (or an abelian) extension; for this reason modding out by
the symmetry group leads to a bundle of projective spaces parametrized 
by the external field configurations.

The obstruction to replacing a projective bundle over $X$ by a true
vector bundle is given by a cohomology class in $H^3(X,\Bbb Z),$ the
\it Dixmier-Douady class. \rm  At the same time, the Dixmier-Douady 
class describes a (stable) equivalence class of a gerbe over X. 
For the general theory of gerbes, the reader is recommended to consult 
[Br]. Here our discussion is closely related to a specialized form,
the \it bundle gerbe, \rm introduced in [Mu], which is an abstraction 
of a quantum field theory problem involving massless fermions, [Mi]. 

A gerbe over $X$ can be viewed as a collection of local line bundles 
$L_{ij}$ over intersections $U_{ij}=U_i\cap U_j$ of open subsets of
$X.$ In addition, the gerbe data involves a family of isomorphisms
$$L_{ij}\otimes L_{jk}= L_{ik}$$ 
on the triple overlaps $U_{ijk}.$ Given a family of curvature forms 
$\omega_{ij}$ for the local line bundles, satisfying the cocycle
property $\omega_{ij}+\omega_{jk}=\omega_{ik},$
one can produce a closed integral 3-form $\omega$ defined on $X$ using 
a standard construction in a Cech - de Rham double complex. The class 
$[\omega] \in H^3(X,\Bbb Z)$ is de Rham form of the Dixmier-Douady
class of the gerbe. (This is not the whole story, because there are 
cases when the DD class is pure torsion.) 

The local line bundles $L_{ij}$ arise in a natural way when trying to
deprojectivisize a projective bundle over $X.$ The transition
functions on the overlaps $U_{ij}$ of a projective bundle are given as
functions $g_{ij}$ taking values in a projective unitary group. The
unitary group $U(H)$ is a central extension by the circle $S^1$ of the
projective unitary group $PU(H).$ Thus the possible unitaries $\hat
g_{ij}(x)$ representing $g_{ij}(x)\in PU(H)$ form a circle over the
point $x\in U_{ij};$ replacing the circle by $\Bbb C$ we obtain a
complex line $L_{ij}(x).$ The group product in $U(H)$ gives a natural
identification of $L_{ij}\otimes L_{jk}$ as $L_{ik}$ in the common 
domain.  

Sections 2 and 3 contain an introduction to the (twisted) K-theory
aspects of gerbes, from the quantum field theory point of view. 
K-theory arises naturally in (hamiltonian) quantization. We have a
family of Hamilton operators parametrized by points in $X.$ When the
Hamilton operators are self-adjoint and have both positive and
negative essential spectrum, they give (by definition) an element of
$K^1(X).$ It is known that $K^1(X)$ is parametrized  
by the odd cohomology groups $H^{2k+1}(X,\Bbb Z).$ The Dixmier-Douady 
class is then the projection to the 3-cohomology part. If we disregard
torsion, we can describe this by a de Rham form of degree 3.   
This is also the starting point for constructing the twisted K-theory 
classes, [Ro].  In section 4 we give an explicit example using the
supersymmetric WZW model. 

Of course, in this short presentation I have left out many interesting
topics; for recent discussions  on  the applications of gerbes and K-theory 
to strings and conformal field theory see e.g. [GR], [Se].

I want to thank Jens Hoppe for pointing out the reference [Ko] and 
Edwin Langmann for critical reading of the manuscript and 
telling me about [B-R]. This work was partially supported by the Erwin 
Schr\"odinger International Institute for Mathematical Physics and by
the  Royal Swedish Academy of Sciences, which is greatfully acknowledged.   
  
\vskip 0.3in

1. GERBES FROM CANONICAL QUANTIZATION 

\vskip 0.3in

Let us first recall some basic facts about canonical anticommutation 
relations (CAR) in infinite dimensions, for an extensive review see
[Ar]. Let $H$ be a complex Hilbert
space. To each vector $u\in H$ one associates a pair of elements 
$a(u)$ and $a^*(u)$ which are generators of a complex unital algebra,
the CAR algebra based on $H.$ The basic relations are 
$$\align &a(u)a^*(v)+ a^*(v) a(u) = \, <u,v>\bold{1}\\
& a(u)a(v)+a(v)a(u)=0=a^*(u)a^*(v)+a^*(v)a^*(u)\tag1.1 \endalign$$
for all vectors $u,v\in H.$ The map $u\mapsto a^*(u)$ is linear
whereas $u\mapsto a(u)$ is antilinear.  

Each polarization $H=H_+\oplus H_-$ to a pair of infinite-dimensional
subspaces defines an irreducible representation of CAR in a dense
domain of a Hilbert space (the \it Fock
space \rm) $\Cal F=  \Cal F(H_+\oplus H_-).$ The (equivalence class of the)
representation is uniquely defined by the requirement of existence of
a \it vacuum vector \rm $\psi_0$ such that 
$$ a^*(u) \psi_0 =0= a(v) \psi_0\tag1.2 $$ 
for all $u\in H_-$ and $v\in H_+.$ Any two representations defined by 
polarizations $H_+, H'_+$ are equivalent if and only if the
projection operators $P,P'$ to these subspaces differ by a
Hilbert-Schmidt operator. 

A unitary operator $S: H\to H$ can be promoted to a unitary operator
$\hat S :\Cal F\to \Cal F$ such that 
$$ \hat S a(u) \hat S^{-1}= a(S u) \,\,\, \forall u\in H,$$
and similarly for the creation operators $a^*(u),$ if and only if 
the off-diagonal blocks of $S$ (in the given polarization) are
Hilbert-Schmidt operators. Let us denote the group of unitaries $S$ of
this type as $U_{res}=U_{res}(H_+\oplus H_-),$ [PS]. 
Note that the operator $\hat S$ is only 
defined up to a phase factor.  The group of quantum operators $\hat S$
forms a central extension $\hat U_{res}$ of $U_{res},$ 
$$1\to S^1\to \hat U_{res} \to U_{res} \to 1.$$
Likewise, a bounded linear operator
$X:H\to H$ can be 'second quantized' as a linear operator $\hat X$ 
in $\Cal F$ such that 
$$[\hat X, a^*(u)]= a^*(Xu),\,\,\, [\hat X, a(u)]= -a(X^* u)\tag1.3 $$ 
for all $u\in H$ if and only if the off-diagonal blocks of $X$ are 
Hilbert-Schmidt. In this case the operator $\hat X$ is uniquely 
defined modulo an additive constant.  The operators $\hat X$ form a
Lie algebra, a central extension of the Lie algebra of the complex
group $GL_{res},$ with commutation relations 
$$[\hat X, \hat Y] =\widehat{[X,Y]} + c(X,Y),$$
where the complex valued 2-cocycle $c$ depends on certain choices
(physically, the choice of normal ordering in the Fock space), but 
its cohomology class is represented by, [Lu], 
$$c(X,Y) = \frac14 \tr\, \epsilon [\epsilon,X][\epsilon ,Y],$$ 
where $\epsilon = P_{H_+} - P_{H_-}.$  

In quantum field theory problems the polarization arises as a
splitting of the 1-particle Hilbert space into positive and negative 
energy subspaces with respect to a (in general unbounded) self adjoint 
Hamilton operator (e.g. a Dirac operator).  

Consider next a case when we have a parametrized family of
Hamilton operators. Let $X$ be some manifold and let for each $x\in X$ a 
self adjoint operator $D_x$ (in a dense domain of) $H$ be given, such
that $D_x$  depends smoothly on the parameter $x$ in some appropriate 
topology.  Tentativly, we would like to construct a family of Fock
spaces $\Cal F_x$ defined by the polarizations $H= H_+(D_x)\oplus
H_-(D_x)$ to positive and negative spectral subspaces. However, in
general this is not possible in a smooth way because of the spectral
flow: each time an eigenvalue $\lambda_n(x)$ of $D_x$ crosses the zero 
mode $\lambda=0$ we have a discontinuity in the polarization and thus
in the construction of the Fock spaces.  

The potential resolution to the above problem lies in the fact that
one is really interested only in the equivalence class of the CAR
representation.  Therefore one is happy with a choice of a function
$x\mapsto P_x,$ where $P_x$ is a projection operator which differs
from the projection operator onto $H_+(D_x)$ by a Hilbert-Schmidt
operator. However, there can be an obstruction to the existence of the
function $P_x$ which depends on the K-theory class of the mapping
$x\mapsto D_x.$ 

Recall the operator theoretic meaning of $K^1(X).$ Let $Fred_*$ be the
space of self-adjoint Fredholm operators in $H$ with both negative and 
positive essential spectrum. Then $K^1(X)$ can be identified as the
space of homotopy classes of maps $X\to Fred_*.$ In particular, a family
$D_x$ of Dirac type operators defines an element in $K^1(X).$ Up to
torsion, $K^1(X)$ is parametrized by the odd de Rham cohomology classes in
$H^*(X,\Bbb Z).$ It turns out that for the existence of the family
$P_x$ only the 3-form part is relevant.

As a concrete example in quantum field theory consider the case of
Dirac operators coupled to vector potentials. Let $\Cal A$ be the space
of $\bold g$ valued 1-forms on a compact odd dimensional spin manifold $M$
where $\bold g$ is the Lie algebra of a compact group $G.$  Each
$A\in\Cal A$ defines a Dirac operator $D_A$ in the space $H$ of square
integrable spinor fields twisted by a representation of $G.$ The 'free'
Dirac operator $D_0$ defines a background polarization $H=H_+\oplus H_-.$
Each potential $A$ and a real number $\lambda$ defines a spectral subspace
$H_+(A,\lambda)$ corresponding to eigenvalues of $D_A$ strictly bigger
than $\lambda.$

Let $Gr_p(H_+)$ be the Grassmann manifold consisting of all closed
subspaces $W\subset H$ such that $P_W -P_{H_+}$ is in the Schatten ideal
$L_p$ of operators $T$ with $|T|^p$ a trace-class operator. One can show
that each $H_+(A,\lambda)$ belongs to $Gr_p(H_+)$ when $p > \text{dim} M.$
All the Grassmannians $Gr_p$ for $p\geq 1$ are homotopy equivalent
and they are also homotopy equivalent to the space $Fred$ of all Fredholm
operators in $H.$ For this reason $Gr_p$ is a classifying space in K-theory.
In particular, the connected components of $Gr_p$ are labelled by the
Fredholm index of the projection $W\to H_+$ and each component is simply
connected. The second integral cohomology (and second homotopy) of each
component is equal to $\Bbb Z.$ For this reason the complex line bundles
are generated by a single element $DET_p.$ In the case $p\leq 2$ the
curvature of the dual line bundle $DET_p^*$ is particularly simple; it
is given as $2\pi$ times  the normalized 2-form
$$ \omega= \frac{1}{16\pi} \tr\, P dP dP.\tag1.4 $$ 

We can cover $\Cal A$ with the open sets $U_{\lambda}=\{A\in \Cal A|
\lambda\notin Spec(D_A)\}.$ On each $U_{\lambda}$ the map $A\mapsto
H_+(A,\lambda)\in Gr_p$ is smooth. For this reason we may pull back the
line bundle $DET_p$ to a line bundle $DET_{p,\lambda}$ over $U_{\lambda}.$
We shall not go into the explicit construction of $DET_p$ here, [MR].
Instead, the difference bundles $DET_{p,\lambda\lambda'}= DET_{p,\lambda}
\otimes DET_{p,\lambda'}^*$ over $U_{\lambda\lambda'}=U_{\lambda}\cap
U_{\lambda'}$ are easy to describe. The fiber of $DET_{p,\lambda\lambda'}$
is simply the top exterior power of the spectral subspace of $D_A$ for
$\lambda < D_A < \lambda'.$ By construction, we have a canonical identification
$$DET_{p,\lambda\lambda'}\otimes DET_{p,\lambda'\lambda''}=DET_{p,\lambda
\lambda''}\tag1.5 $$
on the triple overlaps, for $\lambda<\lambda' <\lambda''.$ We set
$DET_{p,\lambda\lambda'}=DET_{p,\lambda'\lambda}^*$ for $\lambda>\lambda'.$
A system of complex line bundles with the cocycle property (1.5) defines a
a \it gerbe. \rm In this case we have a trivial gerbe, it is generated
by local line bundles $DET_{p,\lambda}$ over the open sets $U_{\lambda}.$
However, we may push things down to the space of gauge orbits $X=\Cal A/
\Cal G,$ where $\Cal G$ is the group of \it based gauge transformations, \rm
i.e., the group of smooth maps $f:M\to G$ such that $f(p)=1$ at a base point
$p\in M;$ here we assume for simplicity that $M$ is connected.

The gauge group acts covariantly on $\Cal A$ and on the eigenvectors of $D_A$
and therefore we may mod out by the $\Cal G$ action to manufacture complex
line bundles over $V_{\lambda\lambda'}= U_{\lambda\lambda'}/\Cal G.$
These line bundles (which we also denote by $DET_{p,\lambda\lambda'}$)
satisfy the same cocycle property (5) as the original bundles over $U_{\lambda
\lambda'}.$  Thus we obtain a gerbe over $X.$ Generically, this gerbe is
nontrivial.  In general, there is an obstruction to the trivialization of the
gerbe given as a \it Dixmier-Douady class \rm in $H^3(X,\Bbb Z).$

The Dixmier-Douady class can be computed as follows. Let $\omega_{\lambda
\lambda'}$ be the curvature form of $DET_{\lambda\lambda'}.$ These satisfy
$$\omega_{\lambda\lambda'} +\omega_{\lambda'\lambda''}=\omega_{\lambda
\lambda''}\tag1.6 $$
on the triple overlaps. Let $\{\rho_{\lambda}\}$ be a partition of unity
subordinate to the covering by the open sets $V_{\lambda}.$ The closed
3-forms
$$\omega_{\lambda}=\sum_{\lambda'} d\rho_{\lambda'} \omega_{\lambda\lambda'}
$$
satisfy $\omega_{\lambda}=\omega_{\lambda'}$ on $V_{\lambda\lambda'}$
and therefore can be glued together to produce a global closed 3-form
$\omega.$ This is easily seen to be integral. The class $[\omega]\in H^3(X,
\Bbb Z)$ is the DD class of the gerbe.

Let us again consider the case of the trivial gerbe over $\Cal A.$ The 
trivialization by the local line bundles $DET_{p,\lambda}$ resolves the
problem of defining a continuous family of CAR representations
parametrized by $\Cal A.$ Over each set $U_{\lambda}$ we can define a
CAR algebra representation in $\Cal F'(A,\lambda)=\Cal F(A,\lambda) 
\otimes DET_{p,\lambda}^*$ by the action $a^*(u)\otimes 1$ and
$a(u)\otimes 1$ on the
fibers. The crux is that the spaces $\Cal F'(A,\lambda)$ are
canonically isomorphic for different values of $\lambda,$ and hence we 
have a well-defined family of spaces  $\Cal F'(A)$ for all $A\in \Cal
A,$ [Mi]. In physics terminology, an isomorphism between $\Cal
F(A,\lambda)$ and $\Cal F(A,\lambda')$ for $\lambda < \lambda'$ is
obtained by 'filling the Dirac sea' from the vacuum level $\lambda$ up 
to the level $\lambda'.$ The filling is canonically defined up to a
unitary rotation of the eigenvectors of $D_A$ in the spectral interval 
$[\lambda,\lambda'];$ a rotation of the basis by $R$  leads to a phase
factor det$R$ in the filling, which is exactly compensated by the
inverse phase factor in the isomorphism between the dual determinant
lines $DET^*_{p,\lambda}(A)$ and $DET^*_{p,\lambda'}(A).$

\vskip 0.3in

2. K-THEORY ASPECTS OF CANONICAL QUANTIZATION\rm

\vskip 0.3in

Let $X$ be a parameter space for a family of self-adjoint operators with
both positive and negative essential spectrum, that is, we have a map
$X\to Fred_*$ or in other words we have an element in $K^1(X).$

As shown in [AS] the space $Fred_*$ is homotopy equivalent to the group
of unitaries $g$ in the complex Hilbert space $H$ such that $g-1$ is compact.
According to Palais this group is homotopy equivalent to the group $U^{(p)}$
of unitaries $g$ such that $g-1\in L_p$ for any $p\geq 1.$ The choice
$G=U^{(1)}$ is the most convenient one since it allows to write in a simple way
the generators for $H^*(G,\Bbb Z).$ The cohomology is generated by
the odd elements
$$c_{2k+1} = N_k \tr\, (dg g^{-1})^{2k+1}$$
for $k=0,1,2,\dots;$ $N_k=-(1/2\pi i)^{k+1} \frac{k!}{(2k+1)!}$ 
is a normalization constant.

The infinite-dimensional  group $U_{res}$  is of interest
to us. Let $\Cal H=L^2(S^1,H).$ Then the group $\Omega G$ of smooth based
loops in $G$ acts naturally in $\Cal H.$ The space $\Cal H$ has a natural
polarization $\Cal H=\Cal H_+\oplus\Cal H_-$ to positive (resp. zero
and negative) Fourier modes. The $\Omega G$ orbit of $\Cal H_+$ lies in
the Hilbert-Schmidt Grassmannian $Gr_2(\Cal H_+).$ Thus  we have
$\Omega G\subset U_{res}(\Cal H_+\oplus \Cal H_-).$ Actually, 
the inclusion is a homotopy equivalence (a consequence of Bott periodicity), [CM1].

There is a universal $\Omega G$ bundle $P$ over $G,$ with total space the
set of smooth paths $f:[0,1]\to G$ starting from the unit element and
such that 
$f^{-1} df$ is  periodic at the end points. Replacing $\Omega G$ by $U_{res}$
we obtain an universal $U_{res}$ bundle over $G.$ Thus $G$ is a classifying
space for $U_{res}$ bundles and we have:

\proclaim{Proposition} $K^1(X)$ is isomorphic to the group of equivalence
classes of $U_{res}$ bundles over $X.$ \endproclaim

The group structure in $K^1(X)$ comes from the representation of
elements in $K^1(X)$ as homotopy classes of maps $g:X\to G.$ The
product is the pointwise multiplication of maps. 

In canonical quantization, it is the $U_{res}$ bundle aspect of $K^1(X)$
which is seen more directly. As we discussed earlier, the Fock representations
of the CAR algebra are determined by polarizations of the 1-particle space.
A family $D_x$ of self-adjoint operators in $Fred_*$ defines a principal
$U_{res}$ bundle over $X.$ The fiber of the bundle at $x$ is the set of unitaries
$g$ in $H=H_+\oplus H_-$  such that the projection onto $g H_+$ differs
from the spectral projection to the subspace $D_x > 0$ by a Hilbert-Schmidt
operator. Since $Fred_*\simeq G$ this bundle is the pull-back of the universal
bundle $P$ over $G$ by the map $x\mapsto D_x.$

The second quantization may be viewed as a prolongation problem for a $U_{res}$
bundle $P_D$ over $X.$ We want to construct a vector bundle $\Cal F$ over
$X$ such that the fibers are Fock spaces carrying representations of the
CAR algebra. The Fock bundle is an associated bundle, not to $P_D$ because
$U_{res}$ does not act in the Fock spaces, but to an $\hat U_{res}$ bundle
$\hat P_D$ which is a prolongation of $P_D$ by the center $S^1$ of
$\hat U_{res}.$ The following was proven in [CM1]:

\proclaim{Theorem 1} There is an obstruction for prolonging $P_D$ to
$\hat P_D$ given by the Dixmier-Douady class which is in the 3-form part
of the K-theory class $[D] \in K^1(X).$ \endproclaim

In many cases the DD class can be computed using index theory, see
[CMM1,2], [CM1,2], but here we shall discuss a bit more the calculation
based on the homotopy equivalence $Fred_*\simeq  G.$

First, one can construct a homotopy equivalence from $Fred_*$ to the space
of \it bounded \rm self-adjoint operators with essential spectrum at
$\pm 1,$ [AS].  For Dirac type operators on a compact manifold we could take
for example the map $D\mapsto F_D= D/(|D| + \exp(-D^2) ).$ This is the
approximate sign operator of $D.$ The next step is to map the $F_D$'s to
unitary operators by $F_D\mapsto g_D= -\exp(i\pi F_D).$ It is not
difficult to see that the difference $g_D -1$ is 
trace-class.  Note that for Dirac operators on a compact manifold of
dimension $n$ we could take $F_D= D/(|D| +1),$ for simplicity, but then
we would have the weaker condition $g_D -1\in L_p$ for $p> n.$

Formally, the pull-back of the universal DD class 
$$c_3= \frac{1}{24\pi^2} \tr\, (dg g^{-1})^3$$
 with respect to the mapping
$F\mapsto g=-\exp(i\pi F)$ can be computed as $\omega= d\theta$ with
$$\theta= \frac18 \tr\, dF h(ad_{i\pi F}) dF\tag2.1 $$
where $h(x)= (\text{sinh}(x) -x)/x^2$ and $ad_F(z)= [F,z].$ However, in
general the expression after 'tr' is not trace-class (and for this reason
$\omega$ is not exact).  There are interesting cases when the above formula
makes sense. When $D$ is a Dirac operator on a manifold of dimension $n$
then one can show using standard estimates on pseudodifferential symbols
that $dF$ is in $L_p$ for any $p >n.$ It follows that the right-hand-side
of (2.1) is well-defined for a Dirac operator on a circle. The case of dimension
$n=3$ is a limiting case. In three dimensions the trace is logarithmically
divergent (when defined as a conditional trace in a basis where $F$ is
diagonal) and subtracting the logarithmic divergence one obtains a finite
expression which can be used to define $\theta.$

The case $F^2=1$ is also interesting; these points define a Grassmann
manifold since $(F+1)/2$ is a projection onto an infinite-dimensional
subspace. Again, in the case of sign operators defined by Dirac
operators one can show that $F$ defines a point in $Gr_p$ for $p>n.$
Assuming that the trace in (2.1) converges, one obtains a specially
simple formula for the form $\theta,$
$$\theta = \frac{1}{16\pi} tr\, F dF dF  \text{ for $F^2=1$ }.$$
Not surprisingly, this is (mod a factor $2\pi$) the curvature formula 
for the determinant bundle $DET_2$ over $Gr_2.$

\vskip 0.3in
3. TWISTED K-THEORY AND QFT 

\vskip 0.3in

There has been an extensive discussion of twisted K-theory in the
recent string theory literature, inspired by suggestions in [Wi]. 
I will not discuss any of the string theory applications
here. Instead, I want to point out that twistings in K-theory are
related to some very basic constructions in standard QFT.

Twisted K-theory was introduced in [Ro] as a generalization of
algebraic K-theory on $C^*$-algebras. Today there are several
equivalent definitions of twisted K, see e.g. [BCMMS]. For QFT 
problems I find it most convenient to use the topologist definition 
of twisted K-theory groups. 
 
Twisted K-theory elements arise from principal $PU(H)$ bundles.  Here
$PU(H)=U(H)/S^1$ is the projective unitary group in a complex Hilbert space.
Since $U(H)$ is contractible by Kuiper's theorem, the homotopy type of
$PU(H)$ is simple: The only nonzero homotopy group is $\pi_2(PU(H))=\Bbb Z.$
For this reason $H^2(PU(H),\Bbb Z)=\Bbb Z.$ On the Lie algebra level, the
basic central extension $1\to S^1\to U(H)\to PU(H)\to 1$ of $PU(H)$ is given
as follows. Let $\psi_0\in H$ be a fixed vector of unit length and
$\bold{g}_0$
the subspace in the Lie algebra $\hat\bold{g}$ of $U(H)$ consisting of
operators
$x$ such that $(\psi_0,x\psi_0)=0.$ For each $y$ in the Lie algebra $\bold g=
\hat\bold{g}/
i\Bbb R$ of $PU(H)$ there is a unique element $\hat y\in \bold{g}_0$ such
that $\pi(\hat y)=y,$ where $\pi$ is the canonical projection.
We can write $\hat\bold{g}=\bold{g} \oplus i\Bbb R$ and the commutation
relations in $\hat \bold{g}$ can be written as
$$[(x,\alpha),(y,\beta)]= ([x,y]_{\bold{g}} , c(x,y) )$$
where the cocycle $c$ is given by $c(x,y)= [\hat x,\hat y] -\widehat{[x,y]}_{
\bold{g}}.$

Given a principal $PU(H)$ bundle $P$ over $X$ we we can construct an associated
$Fred_*$ bundle $Q(P)$ over $X$ as $P\times_{PU(H)} Fred_*$ where $PU(H)$ acts
on $Fred_*$ through conjugation. Again, using the homotopy equivalence $Fred_*
\simeq G$ we might consider $G$ bundles as well; but then it is important
to keep in mind that these are not principal $G$ bundles.
The twisted $K^1$ of $X,$ to be denoted by $K^1(X,[P]),$ is then defined as
the set of homotopy classes of sections of the bundle $Q(P).$ A similar
definition is used for the twisted $K^0$ group, the space $Fred_*$ is then replaced
by the space of all Fredholm operators in $H,$ or alternatively, we can
use the model $U_{res} \simeq Fred.$ 

Quantum field theory provides concrete examples of twisted bundles
$Q(P)$ and its sections.  As we have seen, a family of Dirac type
hamiltonians parametrized by $X$ is an element of $K^1(X)$ or
equivalently, an equivalence class of $U_{res}$ bundles over $X.$ 
The basic observation is that the group $U_{res}$ can be viewed as 
a subgroup of $PU(\Cal F)$ where $\Cal F$ is a Fock space carrying 
a representation of the central extension $\hat U_{res}.$ Therefore 
we can extend the $U_{res}$ bundle to a $PU(\Cal F)$ bundle $P$ over
$X.$ A section of the associated bundle $Q(P)$ becomes now a function
$f$ from $P$ to the space of operators of type $Fred_*$ in the Fock space 
$\Cal F$ such that $f(pg)= g^{-1} f(p) g$ for all $g\in PU(\Cal F)$ 
and $p\in P.$ Since our $PU(\Cal F)$ reduces to a $U_{res}$ bundle 
$P_0$ we might as well construct a section of $Q(P)$ from an
equivariant function on $P_0$ with values in $Fred_*.$

Let $U_i$ be a family of open sets covering $X$ equipped 
with  local trivializations of the $U_{res}$ bundle.  Let $g_{ij}: 
U_i\cap U_j \to U_{res}$ be the corresponding transition functions. 
Then a section of $Q(P)$ can be given by a family of maps 
$$a_i : U_i \to Fred_*(\Cal F)$$
such that
$$ a_j(x) = \hat{g}_{ij}(x)^{-1} a_i(x) \hat{g}_{ij}(x) \,\, 
\text{ for $x\in U_{ij}$ }.$$ 
Here $\hat g$ is the quantum operator, acting in $\Cal F,$
corresponding to the '1-particle operator' $g.$ 

In general, twisted $K^1$ is not easy to compute. Let us consider a
simple example.
 
\bf Example \rm  Let $X=S^3$ which can be identified as the group 
$SU(2)$ of unitary $2\times 2$ matrices of determinant $=1.$ 
A twisted $PU(H)$ bundle $P$ over $S^3$ is constructed as follows. 
First, the space of smooth vector potentials $A$ on a circle with
values in the Lie algebra of $SU(2)$ can be viewed as a principal 
$\Omega SU(2)$ bundle over $S^3.$ The group of based loops $\Omega
SU(2)$ acts on the vector potentials through gauge transformations 
$A\mapsto g^{-1} Ag+ g^{-1}dg.$ A vector potential modulo based gauge 
transformations is parametrized by the holonomy around the circle, 
giving an element in $S^3 =SU(2).$ 

The loop group $\Omega SU(2)$ acts in the space $H$ of
square-integrable  $\Bbb C^2$  valued functions on the unit circle, 
giving an embedding $\Omega SU(2) \subset U_{res}(H_+\oplus H_-),$ 
the polarization being given by the splitting to negative and
nonnegative Fourier modes. Thus we can extend the $\Omega SU(2)$ 
bundle $\Cal A$ over $S^3$ to a principal $U_{res}$ bundle $P_0.$ 
This extends, as explained above, to a principal $PU(\Cal F)$ bundle over 
$S^3.$ 

All principal $PU(\Cal F)$ bundles over $S^3$ are classified by the
homotopy classes of 
transition functions $S^2 \to PU(\Cal F),$ that is, by the the elements 
$n\in \pi_2(PU(\Cal F)) =\Bbb Z.$  The construction above gives the basic 
bundle with $n=1.$ The higher bundles are obtained by taking tensor 
powers (and their duals) of the Fock space representations of the 
central extension of the loop group $\Omega SU(2).$ The
$K^1$-theory twist in this case is fixed by a choice of the integer
$n.$  An element of the twisted $K^1(S^3,n)$ is now given by the
homotopy class of pairs of functions $h_{\pm}: S^3_{\pm}\to Fred_*$ 
such that on the equator $S^2\sim S^3_+\cap S^3_-$ 
$$h_+(x) = g^{-1}(x) h_-(x) g(x),$$
where $g:S^2 \to PU(H)$ is can be given explicitly, using the 
embedding $\Omega SU(2)\subset PU(\Cal F)$   and the fact that 
$\pi_2 \Omega SU(2) =\Bbb Z,$ see [CMM2] for details.  

All classes in $K^1(SU(2),n)$ can actually be given in a simpler way.
We can use the homotopy equivalence $Fred_* \simeq U^{(1)}=G.$ 
Choose then $h_+:S_+^3 \to G$ 
such that it is equal to the unit element on the overlap $S^3_+\cap
S^3_-$ and take $h_-$ as the constant function on $S^3_-$ taking the
value $1\in G.$ Then clearly $h_{\pm}$ are intertwined by the
transition function $g$ on the overlap. Since $h_+$ is constant on the 
boundary of $S^3_+,$ it can be viewed as a map $g_+:S^3 \to G.$  
The winding number of this map in $\pi_3(G)=\Bbb Z,$ modulo $n,$ determines the class in 
$K^1(S^3,n)= \Bbb Z/n\Bbb Z.$ 

The transition function in the present example, being a map from $S^2$ 
to $U_{res}$ (which is a classifying space for $K^0$), 
can also be viewed as an element of $K^0(S^2)= \Bbb Z\oplus\Bbb Z.$ 
This is an essential part of the computation of $K^1(S^3,n),$ based on
the Mayer-Vietoris theorem in K-theory, see [BCMMS] for details, or 
the original computation in [Ro].

\vskip 0.3in
4. AN EXAMPLE: SUPERSYMMETRIC WZW MODEL

\vskip 0.3in
Let $H_b$ be a complex Hilbert space carrying an irreducible unitary
highest weight  
representation of the central extension $\widehat{LG}$ of the loop
group $LG$ of 
level $k;$ here $G$ is assumed to be compact and simple, dim$\, G=N.$  
The level satisfies $2k/\theta^2= 0,1,2,\dots,$ where $\theta$ is the
length of the longest root of $G.$

Let $H_f$ be a fermionic Fock space for the CAR algebra
generated by elements $\psi_n^a$ with $n\in \Bbb Z$ and $a=1,2,\dots, 
N=\text{dim}\, G,$ 
$$\psi^a_n \psi_m^b +\psi_m^b \psi_n^a = 2 \delta_{n,-m}
\delta_{a,b}.\tag4.1$$ 
The Fock vacuum  is a subspace of $H_f$ of dimension $2^{[N/2]}$ (here
$[p]$ denotes the integral part of a real number $p$). The vacuum
subspace carries an irreducible representation of the Clifford algebra
generated by the $\psi^a_0$'s and in addition any vector in the vacuum
subspace is annihilated by all $\psi^a_n$'s with $n<0.$   

The tensor product space $H=H_f \otimes H_b$ carries a tensor product
representation of $\widehat{LG}.$ The fermionic part of the
representation is determined by the requirement 
$$T_f(g) \psi(\alpha) T_f(g)^{-1} = \psi(g   \cdot\alpha),\tag4.2$$ 
where $\alpha$ is a $\Bbb C^N$ valued smooth function on the unit
circle and $\psi(\alpha)= \sum \psi_n^a \alpha_{-n}^a,$ where the 
$\alpha_n^a$'s are the Fourier components of the vector valued 
function $\alpha.$  The action of $g\in LG$ on $\alpha$ is the
point-wise adjoint action on the Lie algebra of the loop group. 

The Lie algebra of $\widehat{LG}$ acting in $H_b$ is generated by the 
Fourier modes $T_n^a$ subject to the commutation relations 
$$[T_n^a,T_m^b ]= \lambda_{abc} T_{n+m}^c + k n\delta_{n,-m}
\delta_{a,b},\tag4.3$$ 
where the $\lambda_{abc}$'s are the structure constants of the Lie
algebra $\bold{g}$ in a basis $T^a$ which is orthonormal with respect
to the Killing form 
$<X,Y> = - \tr(ad_X \cdot ad_Y).$  
There is, up to a
phase factor, a unique normalized vector $x_{\lambda} \in H_b$ such
that $T^a_n x_{\lambda }=0$ for $n<0$ and is a highest weight vector 
of weight $\lambda$ for the finite-dimensional Lie algebra $\bold{g}.$ 

We denote the loop algebra generators acting in the fermionic Fock
space $H_f$ by $K_n^a.$ They satisfy the commutation relations 
$$[K_n^a,K_m^b]= -\lambda_{abc} K^c_{n+m} +\frac12
n\delta_{n,-m}\delta_{a,b}. \tag4.4$$   
Explicitly, the generators are given by
$$K^a_n = -\frac14 \lambda_{abc} :\psi^b_{n+m} \psi^c_{-m}:.\tag4.5$$
The normal ordering $:\, :$ is defined as the rule to write 
the operators with negative momentum index to the right of those with 
positive index. Actually, since our $\lambda_{abc}$'s are totally 
antisymmetric, the normal ordering in (4.5) is irrelevant. 

We denote by $S^a_n$ the generators of the tensor
product representation in $H=H_b\otimes H_f.$ They satisfy the
relations 
$$[S^a_n, S^b_m] = \lambda_{abc} S^c_{n+m} + (k+\frac12)n
\delta_{ab}\delta_{n,-m}.\tag4.6$$

The free hamilton operator is
$$h= h_b\otimes 1 + 1\otimes (2k+1) h_f+ \frac{N}{24} (1\otimes 1)$$ 
where
$$h_b = -: T^a_n T^a_{-n} : \text{ and } h_f = - \frac{1}{4}: n\psi^a_n
\psi^a_{-n}:   \tag4.7$$
We use the conventions 
$$(\psi^a_n)^* = \psi^a_{-n} \text{ and } (T^a_n)^* = -T^a_{-n}.\tag4.8$$  

As seen by a direct computation, the supercharge $Q$ satisfies $Q^2
=h$ and  is defined by 
$$ Q = i\psi^a_{-n} T^a_n  -\frac{i}{12} \lambda_{abc}
\psi^a_n\psi^b_m\psi^c_{-m-n}.\tag4.9$$ 
For a detailed description of the whole super current algebra, see [KT], or
in somewhat different language, [La]. 
Again, by antisymmetry of the structure constants, no normal ordering
is necessary in the last term on the right. 
The general structure of $Q$ has similarities with  Kostant's
cubic Dirac operator, [Ko], (containing a cubic term in the 'gamma matrices' 
$\psi^a_n$); another variant of this operator has been discussed in 
conformal field theory context in [B-R]. Restricting to zero momentum
modes, the operator $Q$ in fact becomes Kostant's operator 
$$\Cal K =  i\gamma^a T^a  - \frac{i}{12} \lambda_{abc}
\gamma^a\gamma^b\gamma^c,\tag4.10$$ 
where $\psi^a_0 =\gamma^a$ are the Euclidean gamma matrices in
dimension $N.$ By the antisymmetry of the structure 
constants the
last term is totally antisymmetrized product of gamma matrices.  
  
The supercharge is a hermitean operator in a dense domain of the Fock
space $H,$ including all the states which are finite linear
combinations of eigenvectors of $h.$  

There is a difference between the cases $N=\text{dim}\,G$ is odd or
even. In the even case we can define a grading operator $\Gamma$ which 
anticommutes with $Q.$ It is given as $\Gamma= (-1)^F \psi^{N+1}_0,$ 
where $F$ is the fermion number operator, $\psi^a_n F+ F \psi^a_n =
\frac{n}{|n|} \psi^a_n$ for $n\neq 0,$ and $\psi^{N+1}_0$ is the
chirality operator on the even dimensional group manifold $G,$ with
eigenvalues $\pm 1.$    

We can couple the supercharge to an external $\bold{g}$ valued vector potential $A$ on
the circle by setting 
$$Q_A = Q +  k \psi^a_n iA^a_{-n}\tag4.11$$ 
where the Fourier components of the Lie algebra valued vector
potential $A$ satisfy $(A^a_n)^*= -A^a_{-n}.$ By a direct computation, 
$$ [S^a_n, Q_A] = k n \psi^a_{n} + k \lambda_{abc} \psi^b_m
A^c_{n-m}.\tag4.12$$ 
This implies that for a finite gauge transformation $f\in LG$ 
$$ S(f) Q_A S(f)^{-1} = Q_{A^f},\tag4.13$$ 
where $A^f = f^{-1} A f +f^{-1} df.$  

\proclaim{Theorem 2} The family $Q_A$ of hermitean operators in $H$
defines an element of the twisted K-theory group $K^1(G,k')$ where the
twist is $k'= (2k+1)/\theta^2$ times the generator of $H^3(G,\Bbb Z).$ \endproclaim 

\demo{Proof} As pointed out in [BCMMS], twisted K-theory classes over $G$
can be thought of as equivariant maps $f: P\to Fred_*,$ where $P$ is
a principal $PU(H)$ bundle over $G$ with a given Dixmier-Douady
invariant $\omega \in H^3(G,\Bbb Z).$ 
The equivariance condition is $f(pg) = g^{-1} f(p)
g $ for $g\in PU(H).$ In the case at hand, the principal bundle $P$ is
obtained by embedding of the loop group $LG \subset PU(H)$ through the projective 
representation of $LG$ of level $k+\frac12.$  As we saw in (4.13), the family
$Q_A$ is equivariant with respect to the (projective) loop group
action.  Finally, the Dixmier-Douady class determined by the level
$k+1/2$ of the projective representation is $k'$ times the generator 
$\frac{1}{24\pi^2} \tr (g^{-1}dg)^3$ on $G= \Cal A/\Omega G.$ 
\enddemo 
 
Note that in the even case the family $Q_A$ gives necessarily a
trivial element in $K^1.$ This follows from the existence of the
operator $\Gamma$ which anticommutes with the hermitean operators
$Q_A.$ Thus there is no net spectral flow for this family of
operators, which is an essential feature in odd K-theory. 

However, in the even case we can define elements in $K^0$ by the 
standard method familiar from the theory of ordinary Dirac operators: 
We can split $Q_A = Q_A^+ + Q_A^-,$ using the chiral projections
$\frac12(\Gamma \pm 1),$ 
where $(Q_A^+)^* = Q_A^-$ is a
pair  of nonhermitean operators with nontrivial index theory. Either 
of the families $Q_A^{\pm}$ can be used to define an element of 
$K^0(G,k').$ Again, we use the observation that elements in
$K^0(G,k')$ can be viewed as equivariant maps from the total space $P$ 
of a principal $PU(H)$ bundle over $G$ to the set $ Fred$ of all Fredholm    
operators in $H.$ 

The operator $Q$ is also of interest in cyclic cohomology.  It can be
used to construct the entire cyclic cocycle of Jaffe, Lesniewski, and 
Osterwalder [JLO] (they considered the case of abelian Wess-Zumino
model). 
The key fact is that the operator $\exp(-sQ^2)$ is
a  trace class operator for any real $s>0;$ in fact, there is an
explicit formula for the trace, it is equal to the product of Kac 
character formulas for two highest weight representations of the loop 
group, one in the bosonic Fock space and the second in the fermionic 
Fock space. 

The second ingredient in cyclic cohomology is an associative algebra 
$\Cal B$ acting in the Hilbert space such that each $[Q,a]$ is a
bounded operator for $a\in\Cal B.$ This is the case for the elements 
$S(f)=f^a_n S^a_{-n}$ in the current algebra for each smooth function $f$ 
on the unit circle. However, the operators $S(f)$ are not
bounded. This should not be a serious problem since the norm of the
restriction of $S(f)$ to a finite energy subspace is growing
polynomially in  energy, whereas $\tr e^{-sQ^2}$ is decreasing 
exponentially in energy. Recall that the even entire JLO cocycle is
composed of terms 
$$\int_{s_i >0, \sum s_i =1} \tr\, \Gamma a_0 e^{-s_0 Q^2} [Q,a_1] e^{-s_1Q^2} 
\dots [Q,a_n] e^{-s_n Q^2} ds_0\dots ds_{n-1}$$ 
with $a_i\in\Cal B.$ This is finite for elements $a_i$ in the current algebra. The above
formula can be used also in the odd case by setting $\Gamma=1.$

Since the twisted K-theory classes above are labelled by the
irreducible highest weight representations of an affine Kac-Moody
algebra, it is natural to ask what is the relation of the twisted
K-theory on $G$ to the Verlinde algebra of $G, $ on a given level $k.$ 
Actually, D. Freed, M. Hopkins and C. Teleman have announced that
there is a product in $K_G(G,k)$ (the $G$ equivariant version of
$K(G,k)$) which makes it isomorphic to the
Verlinde algebra, [F], [FHT]. It would be interesting to understand the
relation of their geometric construction to the algebraic construction 
based on the supersymmetric WZW model. 
   
\vskip 0.3in

REFERENCES

\vskip 0.2in

[AS] M.F. Atiyah and I. Singer: Index theory for skew-adjoint Fredholm 
operators. I.H.E.S. Publ. Math. \bf 37, \rm 305 (1969) 

[Ar] H. Araki: Bogoliubov automorphisms and Fock representations of
canonical anticommutation relations   In: Contemporary Mathematics, vol.
\bf 62, \rm American Mathematical Society, Providence (1987)

[BCMMS] P. Bouwknegt, Alan L. Carey, Varghese Mathai, Michael
K. Murray, and Danny Stevenson: Twisted K-theory and K-theory of
bundle gerbes. hep-th/0106194
 
[Br] J.-L. Brylinski: \it Loop Spaces, Characteristic Classes, and
Geometric Quantization. \rm Birkh\"auser, Boston-Basel-Berlin (1993).

[B-R] L. Brink and P. Ramond: Dirac equations, light cone
supersymmetry, and superconformal algebras. hep-th/9908208.

[CMM1] A.L. Carey, J. Mickelsson, and M.K. Murray: Index theory, gerbes, and
hamiltonian quantization. Commun. Math. Phys. \bf 183, \rm 707 (1997)
      
[CMM2] A.L. Carey, J. Mickelsson, and M.K. Murray: Bundle gerbes
applied to quantum field theory. Rev. Math. Phys. \bf 12, \rm 65
(2000)

[CM1] A.L. Carey and J. Mickelsson: A gerbe obstruction to quantization 
of fermions on odd-dimensional manifolds with
boundary. Lett. Math. Phys. \bf 51, \rm 145 (2000)

[CM2] A.L. Carey and J. Mickelsson: The universal gerbe,
Dixmier-Douady class, and gauge theory. hep-th/0107207

[F] D. Freed: Twisted K-theory and loop groups. math.AT/0206237. 
Publ. in the Proc. of the ICM 2002, Beijing, August 2002. Vol. III, 
p. 419-430 

[FHT] D. Freed, M. Hopkins, and C. Teleman:  Twisted equivariant
 K-theory with complex coefficients. math.AT/0206257

[GR] K. Gawedzki and N. Reis: WZW branes and gerbes. hep-th/0205233 

[JLO]  A. Jaffe,  A. Lesniewski, K. Osterwalder:  Quantum K theory. 1. The Chern character.
 Commun.Math.Phys. \bf 118, \rm 1 (1988) 

[KT] V. Kac and I. Todorov: Superconformal current algebra and their unitary
representations. Commun. Math. Phys. \bf 102, \rm 337 (1985)

[Ko] B. Kostant: A cubic Dirac operator and emergence of Euler
multiplets of representations for equal rank subgroups. Duke Math. J.  
\bf 100, \rm 447 (1999) 

[La] G. Landweber: Multiplicities of representations and Kostants Dirac
operator for equal rank loop groups. Duke Math. Jour. \bf 110, \rm no.1,
121-160 (2001)

[Lu] L.-E. Lundberg: Quasi-free second quantization. Commun. Math. Phys.
\bf 50, \rm 103 (1976)

[Mi] J. Mickelsson: On the hamiltonian approach to commutator anomalies
in $3+1$ dimensions. Phys. Lett. \bf B 241, \rm 70  (1990)

[MR] J. Mickelsson and S. Rajeev: Current algebras in $d+1$ dimensions and
determinant bundles over infinite-dimensional Grassmannians. Commun. Math.
Phys. \bf 116, \rm  365 (1988)

[Mu]  M.K. Murray.: Bundle gerbes.   J. London Math. Soc. (2) \bf 54,
\rm no. 2, 403 (1996)

[PS] A. Pressley and G. Segal: \it Loop Groups. \rm Clarendon Press,
Oxford (1986)

[Ro] J. Rosenberg: Homological invariants of extensions of
$C^*$-algebras. Proc. Symp. in Pure Math. \bf 38, \rm 35 (1982)

[Se] G. Segal: Topological structures in string
theory. Phil. Trans. R. Soc. Lond. A, \bf 359, \rm 1389 (2001) 

[Wi] E. Witten: D-branes and K-theory. J. High Energy Phys. \bf 12, 
\rm 019 (1998); hep-th/9810188  

\enddocument